\documentclass[12pt]{article}

\newcommand{\sect}[1]{\setcounter{equation}{0}\section{#1}}

\newcommand{\eq}{\begin{equation}}
\newcommand{\eqa}{\begin{eqnarray}}
\newcommand{\en}{\end{equation}}
\newcommand{\ena}{\end{eqnarray}}
\newcommand{\enn}{\nonumber \end{equation}}

\def\epsihat{{\widehat{\varepsilon}}}
\def\deltahat{ {\widehat\delta} }
\def\Omhat{\widehat{\Om}}
\def\Vhat{\widehat{V}}
\def\Tau{{\cal{T}}}
\def\Vtildehat{{\widehat{\Vtilde}}}
\def\Rtildehat{{\widehat{\Rtilde}}}
\def\Omtildehat{{\widehat{\Omtilde}}}
\def\ftildehat{\widehat{\ftilde}}
\def\fhat{\widehat{f}}
\def\ghat{\widehat{g}}
\def\phihat{\widehat{\phi}}
\def\dehat{\widehat{\de}}
\def\Ibar{{\bar I}}

\def\ibar{{\bar i}}
\def\Kbar{{\bar K}}

\def\Omtilde{\widetilde{\Om}}
\def\epsitildehat{\widehat{\epsitilde}}

\def\sk{\vskip .4cm}
\def\noi{\noindent}
\def\om{\omega}
\def\al{\alpha}

\def\ga{\gamma}
\def\Ga{\Gamma}

\let \si\sigma
\let \part\partial

\def\unquarto{{1 \over 4}}

\def\unmezzo{{1 \over 2}}
\def\epsi{\varepsilon}
\def\we{\wedge}

\def\de{\delta}

\def\part{\partial}

\def\sk{\vskip .4cm}

\def\noi{\noindent}

\def\X0{X^0}

\def\om{\omega}

\def\al{\alpha}
\def\ga{\gamma}

\def\unquarto{{1 \over 4}}
\def\unmezzo{{1 \over 2}}
\def\epsi{\varepsilon}

\def\we{\wedge}

\def\de{\delta}

\def\Rhat#1#2{ \Rh^{#1}_{~~~#2} }

\def\La{\Lambda}

\def\square{{\,\lower0.9pt\vbox{\hrule \hbox{\vrule height 0.2 cm
\hskip 0.2 cm \vrule height 0.2 cm}\hrule}\,}}

\def\Phi{\phi}

\def\westar{\we_\star}

\def\Vtilde{\widetilde{V}}

\def\Rtilde{\widetilde{R}}

\def\epsitilde{\widetilde{\epsi}}

\def\ftilde{{\tilde f}}

\def\psibar{\bar \psi}

\def\Om{\Omega}

\def\Rhat{\widehat{R}}

\begin{document}

\begin{titlepage}
\begin{center}{\Large \bf Noncommutative gauge fields coupled to noncommutative gravity}
\\[3em]
{\large {\bf Paolo Aschieri} and {\bf Leonardo Castellani}} \\ [2em] {\sl Dipartimento di Scienze e Innovazione Tecnologica
\\ INFN Gruppo collegato di Alessandria,\\Universit\`a del Piemonte Orientale,\\ Viale T. Michel 11,  15121 Alessandria, Italy}\\ [1.5em]
\end{center}

\begin{abstract}

\vskip 0.2cm
We present a noncommutative (NC) version of the action for vielbein
gravity coupled to gauge fields. Noncommutativity is encoded in a twisted $\star$-product between forms, with a set of commuting background vector fields defining the (abelian) twist.  A first order action for the gauge fields avoids the use of the Hodge dual. The NC action is invariant under diffeomorphisms and twisted gauge transformations.  The Seiberg-Witten map, adapted to our geometric setting and generalized for an arbitrary abelian twist, allows to re-express the NC action in terms of classical fields: the result is a deformed action, invariant under  diffeomorphisms and usual gauge transformations. This deformed action is a particular higher derivative extension of the Einstein-Hilbert action coupled to Yang-Mills fields, and to the background vector fields defining the twist. Here noncommutativity of the original NC action dictates the precise form of this extension. We explicitly compute the first order correction in the NC parameter of the deformed action, and find that it is proportional to cubic products of the gauge field strength and to the symmetric anomaly tensor $D_{IJK}$.

 \end{abstract}

\vskip 6.5cm \noi \hrule \vskip.2cm \noi {\small aschieri@to.infn.it\\
leonardo.castellani@mfn.unipmn.it }

\end{titlepage}

\newpage
\setcounter{page}{1}

\sect{Introduction}

We present a noncommutative (NC) version, obtained by a
twist, of the geometric action for gauge fields coupled to
gravity. Via the Seiberg-Witten map \cite{SW}, relating the noncommutative fields to
their classical (commutative) counterparts, we find a higher derivative generalization of Einstein theory coupled to gauge fields, where the higher order couplings are dictated by the noncommutative structure of the original NC action. The resulting theory is geometric (diffeomorphic invariant) and gauge invariant under usual gauge 
transformations.

Thus noncommutativity indicates a particular choice within
the numerous classical extensions of Einstein gravity coupled to Yang-Mills fields. 
In most cases these extensions are seen as effective theories,
with various phenomenological motivations. 

Field theories on noncommutative spaces become especially tractable when the
noncommutativity can be encoded in a twisted product (associative, noncommutative 
$\star$-product) between ordinary fields. 

This product between fields
generates infinitely many derivatives and introduces a
dimensionful noncommutativity parameter $\theta$.  The
prototypical and simplest example of $\star$-product is the
Moyal-Groenewold product \cite{MoyalGroenewold} (historically
arising in phase-space after Weyl quantization \cite{Weyl}) : 
 \eqa & & f(x) \star g(x) \equiv \exp
\left(  {i \over 2} \theta^{\mu\nu}  {\part \over \part x^\mu}
{\part \over \part y^\nu} \right) f(x) g(y) |_{y \rightarrow x} \nonumber \\
& & = f(x) g(x) + {i\over 2} \theta^{\mu \nu}
 \part_\mu f  \part_\nu  g + \cdots + {1 \over n!}  {\left( i \over 2 \right)^n} \theta^{\mu_1\nu_1}
 \cdots \theta^{\mu_n\nu_n} (\part_{\mu_1} \cdots  \part_{\mu_n} f )(\part_{\nu_1} \cdots  \part_{\nu_n} g )+ \cdots \nonumber\\
 & & \label{starproduct}
 \ena
 \noi with a constant $\theta$. An easy generalization  is provided by the twisted $\star$-product, where the partial  derivatives in (\ref{starproduct}) are replaced by a set of commuting
tangent vectors $X_A \equiv  X_A^\mu \part_\mu$. To extend this $\star$-product to forms one simply replaces
the tangent vectors $X_A$, acting on functions, with Lie derivatives along $X_A$, acting
on forms.

 The simple prescription of replacing 
products between fields with $\star$-products yields nonlocal actions, containing
an infinite number of new interactions and higher derivative terms. In this way 
deformed vielbein gravity (coupled to fermions)  \cite{AC1} and supergravity \cite{AC2} have been obtained, as well as
deformed Yang-Mills theories in flat space (see for ex. \cite{Madore,starYM,AJSW}).

These deformed theories, also called noncommutative (NC) theories, are invariant
under deformations of the original symmetries. For example the NC action for gauge fields 
is invariant under deformed gauge symmetries that involve $\star$-products.

Noncommutativity apparently comes with a price, i.e. a proliferation of new degrees
of freedom. This can be simply understood by considering the $\star$-deformation of 
the Yang-Mills field strength:
\eq
F_{\mu\nu}^{I}T_{I} = \part_{\mu} A_{\nu}^{I}T_{I}-\part_{\nu} A_{\mu}^{I}T_{I}-(A_{\mu}^{I} \star A_{\nu}^{J}-
A_{\nu}^{I} \star A_{\mu}^{J})T_{I}T_{J}
\en
Because of noncommutativity of the $\star$-product, anticommutators as well as commutators of
group generators are appearing in the right-hand side, and therefore the $T_{I}$ must  be a basis
for the whole universal enveloping algebra of $G$. The range of the index $I$ increases dramatically,
and so do the number of independent $A^{I}_{\mu}$ field components. A way to reduce this proliferation
is to choose a specific representation for the generators $T_{I}$. For example if the gauge group is
$SU(2)$ and we take its generators to be the in the defining $2 \otimes 2$ representation, these are just
the Pauli matrices, and a basis for the enveloping algebra only requires an additional matrix proportional 
to the unit matrix.
We may even dispense with these remaining extra degrees of freedom if we use the Seiberg-Witten map, that
allows to express the extra fields in terms of the original set of fields of the undeformed theory, the so-called
classical fields. In the $SU(2)$ example, the map relates the four noncommutative fields to 
the three classical $SU{2}$ gauge fields.

With this procedure the NC deformation of vielbein gravity coupled to a fermion field, found in \cite{AC1}, has been
re-expressed in \cite{AC3} in terms of the classical vielbein, spin connection and fermi fields, and its action
has been computed to second order in the noncommutativity parameter.

In the present paper we construct an extension of vielbein gravity coupled to gauge fields, first
by finding a geometric NC action invariant under general coordinate and $\star$-gauge transformations,
and then by solving the SW map for the fields in terms of classical fields. Substituting in the NC action produces the higher-order extension of Einstein gravity coupled to Yang-Mills fields and to the commuting vector fields $X_A$.
Here the $X_A$ fields are nondynamical background fields, but they
can become dynamical with a mechanism similar to the one found in ref.
\cite{ACD}. This is discussed in a forthcoming article \cite{AC4}.

The first order (in $\theta$) correction $S^1$ to the classical action is computed explicitly, and is invariant under diffeomorphisms and ordinary gauge transformations. Moreover, it is proportional to cubic products of the gauge field strength, and to the completely symmetric anomaly tensor $D_{IJK} \equiv Tr(T_I \{T_J,T_K\})$ (this is in agreement with the flat space results of ref. \cite{AJSW}). Thus $S^1$ is nonvanishing only for nonreal representations of $SU(N)$ ($N \neq 2)$. 

The paper is organized as follows: in Section 2 we recall the classical geometric action for gauge fields 
coupled to gravity, where a first order formalism avoids the use of the Hodge dual. Section 3 presents an equivalent 
classical action where gauge potential and spin connection are united  in a single generalized connection.
Its  noncommutative version is obtained in Section 4, where we also discuss its NC symmetries. 
The Seiberg-Witten (SW) map is briefly summarized in Section 5, as well as its geometrical formulation for
a general abelian twist. In Section 6 we apply the SW map to find the first order $\theta$ corrections 
of all the fields and curvatures. Finally, in Section 7 we substitute in the NC action the noncommutative fields by their SW expansions, and find the first order correction to the classical action. Section 8 contains conclusions and outlook. Appendix A contains some elementary notions on  NC geometry based on twist deformations, and Appendix B summarizes our 
gamma matrix conventions.

\sect{Gauge fields coupled to gravity: classical action}

\subsection{Geometric first order action}

We recall the geometric first order action for gauge fields coupled to gravity, see
for example ref. \cite{CDF} (p. 185). This formulation avoids the explicit
introduction of the Hodge dual for the gauge field strength\footnote{an essential feature 
in the group geometric formulation of supergravity theories, see for ex. ref.s \cite{CDF,C92}. }.
For generality we include also a cosmological constant $\Lambda$. The action is:

\eq
 S =  \int  (R^{ab} +\Lambda V^a \we V^b) \we V^c \we V^d \epsilon_{abcd} + Tr( f^{ab} F
 -   {1\over 12} f^{rs} f_{rs} V^a \we V^b  )\we V^c \we V^d \epsilon_{abcd} \label{YMplusgravity}
\en

\noi The fundamental fields are:
 \sk
 \noi i) the one-form spin connection $\om^{ab}$, entering the action through the
Lorentz curvature
 \eq
  R^{ab} \equiv d\om^{ab} - \om^{ac} \we \om^{cb}
  \en
 
\noi ii) the gauge field Lie algebra valued one-form $A$
 \eq
A \equiv A^I_{\mu} T_I dx^{\mu},~~~~~~[T_I,T_J] = C^K_{IJ} T_K: ~G~Lie ~algebra
~generators
 \en
 \noi whose two-form field strength
\eq
 F \equiv F^I T_I \equiv dA - A \we A, ~~~\Longrightarrow F^I = dA^I - C^I_{JK} A^J \we A^K \label{defF}
 \en
 \noi appears in the action. Generators of unitary groups are antihermitian in our conventions. The components $R^{cd}_{~~ab}$ and $F_{ab}^I$ of the 2-forms $R^{cd}$ and $F^I$ 
  are defined by
  \eq
  R^{cd} \equiv R^{cd}_{~~ab}V^a \we V^b, ~~~F^I \equiv F_{ab}^I V^a \we V^b
   \en

 \noi  iii) the zero-form auxiliary antisymmetric field $f_{rs} \equiv f^I_{rs}T_I$;
 \sk
 \noi iv) the vielbein one-form $V^a$
\sk The variation of the action with respect to $f^I_{rs}$
identifies the auxiliary field with the antisymmetric components
of the field strength:
 \eq
  F^I_{ab} = f^I_{ab}
 \en
 \noi The field equation for $\om^{ab}$ gives the zero torsion condition:
  \eq
   d V^a - \om^{ab} \we V^b =0
    \en
     which allows to express the spin connection in terms of derivatives of vielbeins and inverse
     vielbeins (second order formalism).
     
     The variation with respect to the gauge field $A$
 yields the Yang-Mills equations of motion:
 \eq
  D^{G}  (F_{ab}^I V^c \we V^d \epsi_{abcd}) =0 ~~~ \Longrightarrow D_a ^{G \otimes SO(1,3)} F_{ab}^I \equiv 
   \partial_a F_{ab}^I - \om_a^{~ac} f_{cb}^I + \om_a^{bc} f_{ca}^I - C^I_{JK} A^J_a f^K_{ab} =0
   \en
  \noi where as usual curved and flat indices are related via the vielbein and the inverse vielbein ($ \om_a^{bc} = V^\mu_a \om_\mu^{bc},~ \part_a = V^\mu_a  \part_\mu$ etc. ). The zero torsion condition has been used to replace derivatives of vielbein with spin connection terms, and  $D^{G}, D ^{G \otimes SO(1,3)}$ are the $G$-covariant and  $G \otimes SO(1,3)$-covariant exterior derivatives.
  
  Finally varying the vielbein $V^a$ leads to the Einstein equation:
   \eq
    R^{ac}_{~~bc} - \unmezzo \de^a_b R^{cd}_{~~cd} = -(F_{ac}^I F_{bc}^I - \unquarto \de^a_b F_{cd}^I F_{cd}^I )
+ 3 \Lambda \de^a_b
    \en

  \subsection{Index-free action}
    
 The action (\ref{YMplusgravity}) can be recast in index-free form as follows:
\eq 
 S =  \int Tr [i \ga_5 (R \we V \we V+  \Lambda V \we V \we V \we V+  f F \we V \we V+ {1\over 6} f f V \we V \we V \we V)] \label{YMindexfree}
\en
\noi where $R \equiv \unquarto R^{ab} \ga_{ab}$, $f \equiv \unquarto f_{rs}^I \ga_{rs} T_I$,
 $F=F^I T_I$, $V \equiv V^a \ga_a$,
and the trace is taken both on the group $G$ representation and on the
spinor space.
Use of the $D=4$ gamma matrix identities
 \eq
  Tr (\ga_{ab} \ga_c \ga_d \ga_5 ) = -4 i \epsi_{abcd}, ~~~ \ga_{[a} \ga_b \ga_c \ga_{d]} = -i \ga_5 \epsi_{abcd}
   \en
   yields the action (\ref{YMplusgravity}). The index-free definition of the Lorentz curvature is
 \eq
 R = d \om - \om \we \om \label{defR}
 \en
\noi with $\om \equiv \unquarto \om^{ab} \ga_{ab}$. The definitions (\ref{defF}) and (\ref{defR}) imply the
Bianchi identities for $F$ and $R$:
\eq 
 D^G F \equiv dF -A \we F + F \we A =0,~~~D^{SO(1,3)} R \equiv d R - \om \we R + R \we \om =0
 \en

\subsection{Symmetries}

Apart from general coordinate invariance, obtained {\it ab initio}
through the use of Cartan calculus, the action (\ref{YMplusgravity}) is invariant under
gauge and local Lorentz transformations. In index-free form these transformations are
\sk
 1) gauge transformations:
 \eqa
 & & \de^{gauge}_\epsi A=d \epsi - A \epsi + \epsi A ~~~\Rightarrow
\de^{gauge}_\epsi
 F = -F \epsi + \epsi F\\
  & &\de^{gauge}_\epsi V = 0 \\
  & &  \de^{gauge}_\epsi f = - f \epsi + \epsi f\\
 & &  \de^{gauge}_\epsi \om =0 ~~~\Rightarrow
\de^{gauge}_\epsi R = 0
   \ena
   \noi with $\epsi \equiv \epsi^I T_I$;
   \sk
 2) Lorentz transformations:
  \eqa
 & & \de^{Lorentz}_\epsi A= 0 \rightarrow \de^{Lorentz}_\epsi F= 0\\
  & &\de^{Lorentz}_\epsi V =  - V \epsi + \epsi V \label{LorV}\\
  & &  \de^{Lorentz}_\epsi f = -f \epsi + \epsi f \\
 & &  \de^{Lorentz}_\epsi \om =  d \epsi -\om  \epsi + \epsi  \om ~~~\Rightarrow
\de^{gauge}_\epsi  R= -R \epsi + \epsi R \label{Lorom}
   \ena
\noi with 
 $ \epsi \equiv \unquarto \epsi^{ab} \ga_{ab}$. Checking the invariance of the action 
 (\ref{YMindexfree}) under these transformations is immediate, using the cyclicity of the trace and the fact that $\epsi$ 
commutes with $\ga_5$ both for gauge and Lorentz transformations.

\section{An improved classical action}

The similar role played by $R$ and $F$ in the action (\ref{YMindexfree}) suggests to unify gauge potential and spin connection in a ``total"  connection $\Om$:
 \eq
 \Om \equiv \om + A = \unquarto \om^{ab} \ga_{ab} + A^I T_I
\en
 and define a total curvature $R(\Om)$ as
 \eq
  R(\Om) \equiv d \Om - \Om \we \Om 
 \en
satisfying  the Bianchi identity:
 \eq
 D R(\Om)  \equiv  d R (\Om) - \Om \we R (\Om) +  R (\Om)  \we \Om =0
 \en
(equivalent to the Bianchi identities for $F$ and $R$ of previous Section), $D$ being the $G \otimes SO(1,3)$-covariant exterior derivative.
  Since $\ga$  matrices commute with the $G$ group generators (they act on different spaces)
 the total curvature is the sum of $R$ and $F$:
 \eq
 R(\Om) = d \om + dA - \om \we \om - A \we A   = R + F  =  \unquarto R^{ab} \ga_{ab} + F^I T_I
 \label{ROmega}
 \en
 We can now consider the following ``improved" index-free action:
  \eq 
 S =  \int Tr [i \ga_5 ( f
R(\Om)  \we V \we V +  {1\over 6} f f~ V \we V \we V \we V + \Lambda' V \we V \we V \we V)] \label{YMindexfreeimproved}
\en
 where the auxiliary 0-form field $f$ is defined as
 \eq
 f \equiv \unquarto f_{rs}^I \ga_{rs} T_I + 1 \label{fimproved}
 \en

It is easy to see that by substituting the expressions for $R(\Om)$ and $f$ given in
 (\ref{ROmega}) and (\ref{fimproved}), and provided that $Tr(T_I)=0$ (holding for any representation and for any product of simple Lie groups\footnote{$Tr(T_I)$ does not necessarily hold for $U(1)$, but choosing the only generator $T$ to be in the defining representation of $SO(2)$ still implies $Tr(T)=0$}),  this action reproduces the index-free action (\ref{YMindexfree}) 
 with a cosmological constant $\Lambda$ given by
 \eq
 \Lambda = \Lambda' + {1 \over 6}
 \en
 Thus $\La' = -(1/6)$ corresponds to a vanishing cosmological constant $\La$.
The improved action is invariant under the total gauge symmetry:
 \eqa
 & & \de_\epsi \Om=d \epsi - \Om \epsi + \epsi \Om ~~~ \Longrightarrow
\de_\epsi
 R(\Om) = -R(\Om) \epsi + \epsi R(\Om) \label{generalgauge}\\
 & &\de_\epsi V =  - V \epsi + \epsi V \label{generalV}\\
  & &  \de_\epsi f = - f \epsi + \epsi f \label{generalf}
   \ena
   \noi  where $\epsi$ depends on spacetime, commutes with $\ga_5$  and otherwise  has an arbitrary structure.  Lorentz and gauge transformations correspond to a parameter
   \eq
  \epsi \equiv \unquarto \epsi^{ab} \ga_{ab} + \epsi^I T_I \label{gaugeLorparameter}
   \en
  \noi Since $T_I$ and $ \ga_{ab}$ commute, 
   the spin connection and the vielbein are inert under gauge transformations,
  and transform under Lorentz transformations as in (\ref{LorV}), (\ref{Lorom}), or in components:
   \eq
     \de^{Lorentz}_\epsi \om^{ab}=d \epsi^{ab}  + \unmezzo \epsi^{ac} \om^{cb} - \unmezzo \epsi^{bc }\om^{ca},~~~~ \de^{Lorentz}_\epsi V^a=  \epsi^{ab} V^b  
      \en
  The gaugel transformations (\ref{generalgauge})-(\ref{generalf}) close according to
 the commutation rules
 \eq
  [\de_{\epsi_1}, \de_{\epsi_2}] = \de_{\epsi_2 \epsi_1- \epsi_1\epsi_2}
   \en
   and for the specific parameter (\ref{gaugeLorparameter}) reproduce the commutation relations of infinitesimal $SO(1,3) \otimes G $ transformations. 

\sect{Noncommutative action}

The twisted noncommutative action is found by replacing in the improved 
index-free action (\ref{YMindexfreeimproved})  all products by twisted
$\star$-products:
\eq
 S =  \int Tr [i \ga_5 ( \fhat \star \Vhat \westar
\Rhat (\Omhat) \westar \Vhat +   {1\over 6} ~ \fhat \star \fhat \star \Vhat \westar
\Vhat  \westar \Vhat \westar \Vhat + 
\Lambda'  \Vhat  \westar \Vhat \westar \Vhat
 \westar \Vhat)] \label{YMtwisted}
\en
\noi and  the  curvature $\Rhat (\Omhat)$ is defined by
 \eq
\Rhat (\Omhat) = d \Omhat- \Omhat \westar \Omhat
 \en
This definition implies the Bianchi identity:
 \eq
 D \Rhat(\Omhat) \equiv  d \Rhat(\Omhat) - \Omhat \westar \Rhat(\Omhat) +  \Rhat(\Omhat)  \westar \Omhat =0
 \en

The particular order $fVRV$ in the first term of the action ensures its reality, as discussed later.
The (associative) $\star$-exterior product between forms is defined by using Lie derivatives along a set
of commuting vector fields $X_A$, see Appendix A. 

The fields in the above NC action have deformed transformation laws: to distinguish them from the ordinary fields transforming under the usual laws we denote them with a hat. In fact, the Seiberg-Witten map 
relates the hatted fields (the "noncommutative" fields) to the ordinary ones.

\subsection{Noncommutative symmetries}

 The NC action (\ref{YMtwisted}) is invariant under general coordinate transformations (being the integral of a 
 4-form) and under the $\star$-gauge variations:
\eqa
 {\deltahat_\epsihat}
 \Omhat&=&d \epsihat - \Omhat \star \epsihat + \epsihat \star \Omhat ~~\Longrightarrow
{\deltahat_\epsihat}
 \Rhat(\Omhat) = -\Rhat(\Omhat)  \star \epsihat + \epsihat \star
 \Rhat(\Omhat) \\
  \deltahat_\epsihat \Vhat &=& -\Vhat \star \epsihat + \epsihat \star
  \Vhat  \\
 \deltahat_\epsihat \fhat &=& - \fhat \star \epsihat + \epsihat \star \fhat
   \ena
   with an arbitrary parameter $\epsihat$ commuting with $\ga_5$. 

The invariance of the noncommutative action under these transformations
relies on the cyclicity of the integral (and of the trace) and on
the fact that $\epsi$ commutes with $\ga_5$.

\subsection{How many fields ?}

Because of noncommutativity, extra fields are entering in
the expansions of $\Omhat, \Vhat, \fhat$. Indeed the set of gauge group generators $\Tau_B=\{T_I,\gamma_{ab}\}$ must include now
all products $\Tau_B \Tau_C$ since both commutators and anticommutators are involved in expressions
like $\Omhat \westar \Omhat = \Omhat^B \westar \Omhat^C\Tau_B \Tau_C$. 
Thus the gauge field, its
curvature and the scalar fields $\widehat f$ take 
values in $U( so(1,3) \otimes g)$, the universal enveloping algebra of $ SO(3,1) \times G $, whose generators are $i \ga_{ab} \otimes T_\Ibar $, $1 \otimes T_\Ibar$ and $i \ga_5 \otimes T_\Ibar$, where $T_\Ibar$ now stand for all (symmetrized) products
of the original $G$ Lie algebra generators $T_I$  plus the unity (conventionally
indicated as $T_0$).
Similarly the noncommutative vielbein $\Vhat$, which  classically has
only $\ga_a$ components, now acquires components along $\ga_a \otimes
T_\Ibar$ and $\ga_a \ga_5 \otimes T_\Ibar$ (the terms proportional to
$\ga_a\ga_5$ are needed because
$\{\gamma_{ab},\gamma_c\}=2i\epsi_{abcd}\gamma^d\gamma_5$, we therefore
  have to consider all odd gamma matrices). Since there are infinite  $T_\Ibar$ generators of the universal enveloping algebra, we would end up with an infinite number of field components.

 However the situation changes drastically if we choose a specific $G$ (irreducible) representation for the $T_\Ibar$ generators: the $U(g)$ enveloping algebra becomes then finite dimensional. It also explicitly includes the unit matrix, since for example the quadratic Casimir invariant for $G$ must be proportional to unity. All generators $T_\Ibar$ are chosen to be antihermitian in our conventions. The antihermitian unity will be written as $T_0 \equiv i 1$. For example if $G=SU(2)$ we can choose the $2 \times 2$ defining representation, and a basis of generators $T_\Ibar$ for the enveloping algebra is given by the four matrices  ($\tau_i=i\sigma_i$, $i1$), where $\sigma_i$ are the Pauli matrices.

The $SO(1,3)$ enveloping algebra generators  $\Ga_\al$
are already in the spinor representation given by Dirac matrices, and are chosen to be $\ga_0$-antihermitian:
 \eqa
 & & \Ga_\al  = {1\over 4} \ga_{ab},~i1,~\ga_5 \\
 & & (\Ga_\al)^\dagger = - \ga_0 \Ga_\al \ga_0
\ena

The noncommutative fields are expanded as follows:
 \eqa
 & & \Omhat = -{i \over 4} \Omhat^{ab,\Ibar}  \ga_{ab} T_\Ibar + \Omhat^\Ibar T_\Ibar -i  \Omtildehat{}^\Ibar \ga_5 T_\Ibar  \label{omegafield}\\
 & & \Vhat = -i \Vhat^{a,\Ibar} \ga_a T_\Ibar - i \Vtildehat{}^{a,\Ibar} \ga_a \ga_5 T_\Ibar\\
 & & \fhat = {1\over 4} \fhat^\Ibar_{ab} \ga^{ab}T_\Ibar +  i \fhat^\Ibar T_\Ibar + 
  \ftildehat{}^\Ibar \ga_5 T_\Ibar \label{fexpansion}
 \ena
where we recall that the genertors $T_{\bar I}$ are
antihermitian. 
Similarly for the curvature and the gauge parameter the expansions are: 
 \eqa
& &
 \Rhat= -{i \over 4} \Rhat^{ab,\Ibar}  \ga_{ab} T_\Ibar + \Rhat^\Ibar T_\Ibar -i
\Rtildehat{}^\Ibar \ga_5 T_\Ibar  \\
& & \epsihat =  -{i \over 4} \epsihat^{ab,\Ibar}  \ga_{ab} T_\Ibar + \epsihat^\Ibar T_\Ibar 
-i \epsitildehat{}^\Ibar \ga_5 T_\Ibar
\ena
 All the components along the $SO(1,3) \times G$ enveloping algebra generators are taken to be real, and therefore fields and curvatures satisfy the hermiticity properties: 
 \eq
 \Omhat^\dagger = - \ga_0 \Omhat \ga_0,~\Vhat^\dagger = \ga_0 \Vhat \ga_0,~\fhat^\dagger = \ga_0 \fhat \ga_0,~\Rhat^\dagger =- \ga_0 \Rhat \ga_0
 \en 
\noi i.e. $\Omhat$ and $\Rhat$ are $\ga_0$-antihermitian, while $V$ and $f$ are $\ga_0$-hermitian. Using these rules it is a quick matter to check that the noncommutative action
(\ref{YMtwisted}) is real. 

\noi In the following we will  occasionally split the index $ \Ibar$ as $T_\Ibar = (T_0 \equiv i1, T_\ibar)$.

We can insist and require that only the components 
$\Vhat^{a,0}$, $ \Omhat^{ab,0}$,  $\Omhat^I$ and $\fhat^I_{rs}$ of the fields given above are independent, in order to match the degrees of
freedom of the classical theory. This is done in Section 5 via the Seiberg-Witten map, that allows  
 to express all noncommutative fields in terms of the 
commutative -or classical- fields $V^a,~\omega^{ab},~ A^I,~ f^I_{rs}~.$
\sk

\section{Seiberg-Witten map and fields at first order in  $\theta$ for Moyal $\star$-product}

The Seiberg-Witten map relates the noncommutative gauge field $\Omhat$
to the ordinary one $\Om$, and the noncommutative gauge parameter
$\epsihat$ to  the ordinary one $\epsi$  so as to 
satisfy:
 \eq
 \Omhat (\Om) + {\widehat{\delta}}_\epsihat
\Omhat (\Om) = \Omhat (\Om + \de_\epsi \Om)    
 \label{SWcondition}
 \en
 with 
  \eqa 
   & &
  \de_\epsi \Om_\mu = \part_\mu \epsi + \epsi \Om_\mu -  \Om_\mu 
      \epsi, \\
      & &
  \deltahat_\epsihat{} \Omhat_\mu = \part_\mu \epsihat + \epsihat \star \Omhat_\mu -  \Omhat_\mu \star     
      \epsihat
      \ena
   
\noi  In words: the dependence of the noncommutative gauge field on the ordinary one is fixed
by requiring that ordinary gauge variations of $\Om$ inside $\Omhat(\Om)$ produce the noncommutative
gauge variation of $\Omhat$.  The above equation can be solved order by order in $\theta$, yielding
$\Omhat$ and $\epsihat$ as a power series in $\theta$:  
 \eqa 
    \Omhat (\Om, \theta) &=& \Om + \Om^1 (\Om)  + \Om^2 (\Om) + \cdots + \Om^n (\Om)+ \cdots  \\
     \epsihat (\epsi, \Om, \theta)  &=&  \epsi + \epsi^1 (\epsi, \Om)+ \epsi^2 (\epsi, \Om)+ \cdots + 
     \epsi^n (\epsi, \Om)+ \cdots 
     \ena    
 \noi  where $\Om^n (\Om)$ and $\epsi^n (\epsi, \Om)$  are of order $n$ in $\theta$. Note that  $\epsihat$
depends on the ordinary $\epsi$ and also on $\Om$.

The recurrence relations that solve (\ref{SWcondition}), in our antihermitian fields convention, are (see ref.  \cite{Ulker},  or \cite{AC3} for a simple proof).
\eqa
& & 
\Omega_\mu^{n+1}=\frac{i}{4(n+1)}\theta^{\rho\sigma}\{\hat\Omega_\rho
, \partial_\sigma \hat\Omega_\mu+2 \hat R(\Omega)_{\rho\mu}\}^n_\star \label{Omn1}\\
& & \epsi^{n+1}=  {i \over 4(n+1)} \theta^{\rho\sigma} \{\Omhat_\rho, \part_\sigma \epsihat \}^n_\star \label{epsin1}
 \ena
   \noi where  $ \{ \fhat, \ghat \}^n_\star $ is $n$-th  order term in $ \{ \fhat, \ghat \}_\star  \equiv f \star g + g \star f $, so that for example
     \eq
     \{\Omhat_\rho, \part_\sigma \epsihat \}^n_\star \equiv \sum_{r+s+t=n} (\Om^r_\rho \star^s \part_\sigma \epsi^t +
      \part_\sigma \epsi^t \star^s \Om^r_\rho )
       \en
    \noi and $\star^s$ indicates the $s$-th order term in the star
    product expansion (\ref{starproduct}). 

Similar expressions hold for the gauge field strength $\Rhat (\Omhat)$ , and for matter fields $\phi$  transforming
 in the adjoint representation of the gauge group:
\eqa
 & & R^{n+1}_{\mu\nu}  = {i \over 4(n+1)} \theta^{\rho\sigma} \left( \{\Omhat_\rho, \part_\sigma \Rhat_{\mu\nu}   
     + D_\sigma \Rhat_{\mu\nu} \}^n_\star - 4\{\Rhat_{\mu\rho},\Rhat_{\nu\sigma} \}^n_\star \right) \label{Rn1}\\
 & & \phi^{n+1}= {i \over 4(n+1)} \theta^{\rho\sigma} \{\Omhat_\rho, \part_\sigma \phihat
     + D_\sigma \phihat \}^n_\star, ~~\dehat_\epsihat \phihat =  \epsihat \star \phihat -  \phihat \star \epsihat 
\label{phin1}
\ena
where the covariant derivatives on $\Rhat$ and $\phihat$  are given by
$D_\sigma\Rhat_{\mu\nu}=\partial_\sigma\Rhat_{\mu\nu}-\Omhat_\sigma\star
\Rhat_{\mu\nu}+ \Rhat_{\mu\nu}\star\Omhat_\sigma$ and
$D_\sigma \phihat = \part_\sigma \phihat - \Omhat_\si \star \phihat +  \phihat \star \Omhat_\si$.
\sk

At first order in $\theta$, recalling that the classical gauge field is $\Om = \om + A =
\unquarto \om^{ab} \ga_{ab} + A^I T_I $ we find the component fields
as collected in the following Table.
\vfill\eject

\noi {\bf TABLE 1: fields at first order in $\theta$}
\sk
\noi  {\bf Spin connection}
 \eqa
 & & \Om^{1~ab,I}_{\mu}=-{1\over 4}\theta^{\rho\sigma} [\om^{ab}_{\rho} (\part_{\sigma}
A^{I}_{\mu} + F^{I}_{\sigma\mu}) + A^{I}_{\rho} (\part_{\sigma}\om^{ab}_{\mu}+R^{ab}_{\si\mu})] \\
 & & \Om^{1~0}_{\mu}=-{1\over 16}\theta^{\rho\sigma} [\om^{ab}_{\rho} (\part_{\sigma}\om^{ab}_{\mu}+  R^{ab}_{\si\mu}) + 8 A^{I}_{\rho} (\part_{\sigma} A^{J}_{\mu} + F^{J}_{\sigma\mu})K^0_{IJ}] \\
 & &   \Om^{1~\ibar}_{\mu}=-{1\over 4}  \theta^{\rho\sigma}  A^{I}_{\rho} (\part_{\sigma} A^{J}_{\mu} + F^{J}_{\sigma\mu}) K^\ibar_{IJ} \\
 & &   \Omtilde^{1~I}_{\mu}= {1 \over 32}  \theta^{\rho\sigma}   \om^{ab}_{\rho} (\part_{\sigma}\om^{cd}_{\mu}+  R^{cd}_{\si\mu})
  \epsilon_{abcd} \de^{I0}
 \ena
where $\{T_I,T_J\}=iK^\Kbar_{IJ}T_\Kbar$, with $K^\Kbar_{IJ}$ real constants.
\sk
\noi {\bf Vielbein}
\eqa
 & & V^{1~a,I}_{\mu} = -\unmezzo \theta^{\rho\sigma} A^I_{\rho}(\part_{\sigma}+ D_\si) V^{a}_{\mu} \\
 & & \Vtilde^{1~a,I}_\mu=-{1 \over 8}  \theta^{\rho \sigma} \om^{bc}_{\rho} \epsilon_{bcda}
     (\part_{\sigma} + D_\si ) V^{d}_{\mu} \de^{I0} 
\ena

\noi {\bf Auxiliary field}
 
 \eqa
 & & f^{1~ab,\Ibar} =- {1\over 4} \theta^{\rho\si} A^J_\rho (\part_\si + D_\si) f^{ab,K} K^\Ibar_{JK}  \\
  & & f^{1~I} = - {1 \over 16} \theta^{\rho\si} \om^{ab}_\rho (\part_\si + D_\si) f^{ab,I}  \\
    & & \ftilde^{1~I} =  {1 \over 32} \theta^{\rho\si} \om^{ab}_\rho (\part_\si + D_\si) f^{cd,I}  \epsilon_{abcd} 
    \ena
  
  \noi {\bf Curvature}
  
  \eqa
  & & R^{1~ab,I}_{\mu\nu} = \theta^{\rho\si} [ - \unmezzo \om^{ab}_\rho (\part_\si + D_\si) F^I_{\mu\nu} - \unmezzo A^I_\rho 
  (\part_\si + D_\si) R^{ab}_{\mu\nu} + R^{ab}_{\mu\rho} F^I_{\nu\si} + F^I_{\mu\rho} R^{ab}_{\nu\si}] \nonumber \\
  & & \\
  & & R^{1~0}_{\mu\nu} =- {1 \over 4} \theta^{\rho\si} [  {1 \over 4}  \om^{ab}_\rho (\part_\si + D_\si) R^{ab}_{\mu\nu}
   - A^I_\rho (\part_\si + D_\si) F^J_{\mu\nu}  K^0_{IJ} - \unmezzo R^{ab}_{\mu\rho} R^{ab}_{\nu\si} + 
   2 F^I_{\mu\rho} F^J_{\nu\si}  K^0_{IJ}] \nonumber\\
    & & \\
   & & R^{1~\ibar}_{\mu\nu} = -{1 \over 4} \theta^{\rho\si} [ 
    A^I_\rho (\part_\si + D_\si) F^J_{\mu\nu}  K^\ibar_{IJ} -  2 F^I_{\mu\rho} F^J_{\nu\si}  K^\ibar_{IJ}] \\
     & & \Rtilde^{1~0}_{\mu\nu} = {1 \over 4} \theta^{\rho\si} [  {1 \over 8}  \om^{ab}_\rho (\part_\si + D_\si) R^{cd}_{\mu\nu}
      \epsilon_{abcd} 
    - \unquarto R^{ab}_{\mu\rho} R^{cd}_{\nu\si} \epsilon_{abcd} ] \\      
    & &\Rtilde^{1~\ibar}_{\mu\nu}  =0
  \ena

\section{Geometric Seiberg-Witten map and fields at first order in $\theta$ for a general abelian twist}

As observed in ref. \cite{AC3}, the SW map can be recast in a coordinate-independent way, and generalized to
a $\star$-product originating from an abelian twist. We recall here the relevant formulae, corresponding to  (\ref{Omn1}), (\ref{epsin1}), (\ref{Rn1}), (\ref{phin1}) of the previous Section:

\eqa
& & 
\Omega^{n+1}=\frac{i}{4(n+1)}\theta^{AB}\{\hat\Omega_A
, \ell_B \hat\Omega+ \hat R_{B} \}^n_\star\\
& & \epsi^{n+1}=  {i \over 4(n+1)} \theta^{AB} \{\Omhat_A, \ell_B \epsihat \}^n_\star \\
 & & R^{n+1} = {i \over 4(n+1)} \theta^{AB} \left( \{\Omhat_A, (\ell_B + L_B)  \Rhat  \}^n_\star 
 - [\Rhat_{A},\Rhat_{B} ]^n_\star \right) \\
 & & \phi^{n+1}= {i \over 4(n+1)} \theta^{AB} \{\Omhat_A,  (\ell_B + L_B) \phihat \}^n_\star,
  ~~\dehat_\epsihat \phihat =  \epsihat \star \phihat -  \phihat \star \epsihat 
\ena
where
$\Omhat_A$, $\Rhat_A$ are defined as the contraction along the tangent
vector $X_A$ of
the exterior forms $\Omhat$, $\Rhat$, i.e. $\Omhat_A\equiv i_A\Omhat$,
$\Rhat_A \equiv i_A \Rhat$, ($i_A$ being the contraction along $X_A$).
We have also intoduced the covariant Lie derivative $L_B$ along the
tangent vector $X_B$; it acts on $\Rhat$ and $\phihat$ as 
$L_B \Rhat =\ell_B \Rhat-\Omhat_B \star
\Rhat+ \Rhat \star\Omhat_B$ and
$L_B  \phihat = \ell_B \phihat - \Omhat_B \star \phihat +  \phihat \star \Omhat_B$.
In fact the covariant Lie derivative $L_B$ has the Cartan form:
 \eq
  L_B = i_B D + D i_B~
    \en
where $D$ is the covariant derivative.

For the fields of the index-free geometrical action (\ref{YMtwisted}) the above formulae at first order in $\theta$ become:
 \eqa
& & f^{1}= {i \over 4} \theta^{AB} \{\Om_A,  (\ell_B + L_B) f \} \label{f1} \\
 & & V^1 = {i \over 4} \theta^{AB} \{\Om_A,  (\ell_B + L_B) V \} \label{V1} \\
 & & R^{1} = {i \over 4} \theta^{AB} \left( \{\Om_A, (\ell_B + L_B)  R  \}
 - [R_{A},R_{B} ] \right) \label{R1}
 \ena
 All these formulae are {\it not} $SO(1,3) \otimes G$-gauge covariant, due to the presence of the ``naked" connection
$\Omhat$ and the non-covariant Lie derivative $\ell_A$. However, when inserted in the NC action (\ref{YMtwisted}), the resulting action is gauge invariant order by order. Indeed usual gauge variations induce the $\star$-gauge variations under which the NC action is invariant. Therefore the NC action, re-expressed in terms of ordinary fields via the SW map, is invariant under usual gauge transformations. Since these do not involve $\theta$, the expanded action is invariant under ordinary gauge variations order by order in $\theta$. This will be explicitly cheked in next Section for the first order $\theta$ correction 
of the NC action. 

\sect{Action at first order in $\theta$}

The first order correction of the action (\ref{YMtwisted}) reads:
 \eqa
  S^1 =& & \int Tr[i \ga_5 ( f^1 (V \we R \we V) + f (V \westar R \westar V)^1 + \nonumber \\
 & &  + {1 \over 6} (f \star f)^1 V \we V \we V \we V + {1 \over 6} ff (V \westar V \westar V \westar V)^1 + \nonumber \\
 & & + 2 \Lambda' (V \westar V)^1 \we V \we V)] \equiv \nonumber \\
 & & \equiv S^1_{fVRV} + S^1_{ffVVVV} + S^1_{VVVV}
 \ena
where one of the $\star$-products in the integrands of (\ref{YMtwisted}) has been replaced by an ordinary product
after partial integration. 
We now insert the first order expressions (\ref{f1}), (\ref{V1}) and (\ref{R1}) and compute separately the three lines, called $S^1_{fVRV}$, $S^1_{ffVVVV}$ and   $S^1_{VVVV}$:
\eqa
 & & S^1_{fVRV}=- \unmezzo \theta^{AB} \int Tr [\ga_5 ((R_{AB} f + fR_{AB})V \we R \we V  \nonumber  \\
 & &~~~~~~~~~~~~ +  f ( L_A V \we  L_B R \we V + L_A V \we R \we L_B V + V \we L_A R \we L_B V ) \nonumber \\
 & & ~~~~~~~~~~~~ - \unmezzo f V \we (R_A \we R_B - R_B \we R_A )  \we V)]
  \ena
 \eqa
 & & S^1_{ffVVVV} = -{1 \over 12} \theta^{AB} \int Tr [\ga_5 ((R_{AB}ff + ffR_{AB}) V \we V \we V \we V \nonumber \\
 & & ~~~~~~~~~~~~ + ff ( L_A V \we L_B V \we V \we V + L_A V \we V \we L_B V \we V + L_A V \we V \we V \we L_B V \nonumber \\
 & & ~~~~~~~~~~~~ + V \we L_A V \we L_B V \we V + V \we L_A V \we V \we L_B V + V \we V \we L_A V \we L_B V) \nonumber \\
 & & ~~~~~~~~~~~~ + (L_A f)(L_B f) V \we V \we V \we V )] \nonumber \\
 \ena
 \eq
 S^1_{VVVV} = - \theta^{AB} \int Tr [\ga_5 \La' (R_{AB}V \we V \we V \we V + L_A V \we L_B V \we V \we V)]
 \en
 where all fields are classical. All three terms are separately gauge invariant. 
  Covariant Lie derivatives 
can be expressed in terms of covariant exterior derivatives through the identities:
 \eqa
 & & L_A V = D i_A V + i_a (D V) = DV_A + i_A T, ~~~V_A \equiv i_A V = X_A^\mu V_\mu\\
 & & L_A R = D R_A \\
 & & L_A f = D_A f 
 \ena
\noi where the torsion $T$ is defined by $T = DV  \equiv dV - \om \we V - V \we \om$. Moreover the curvature components
$R_{AB}^{rs}$ are defined as:
 \eqa
 R_{AB}^{rs} &\equiv &\unmezzo (\ell_A \om_B^{rs} - \ell_B \om_A^{rs} - \om_A^{rt} \we \om_B^{ts} + \om_B^{rt} \we \om_A^{ts})\nonumber  \\
&   = &\unmezzo X_A^\mu X_B^\nu (\part_\mu \om_\nu^{rs} - \part_\nu \om_\mu^{rs} - \om_\mu^{rt} \we \om_\nu^{ts}
+ \om_\nu^{rt} \we \om_\mu^{ts}) =  X_A^\mu X_B^\nu
R_{\mu\nu}^{rs}\nonumber\\
&=&-{\frac{1}{2}}i_Ai_B R^{rs}
 \ena
 \noi the equality between first and second line holding because the vector fields $X_A$ are commuting:
 \eq
 [X_A,X_B] =0 \Rightarrow X_A^\mu \part_\mu X_B^\nu - X_B^\mu \part_\mu X_A^\nu =0
 \en

 Finally, using the classical expressions 
\eq
f \equiv \unquarto f_{rs}^I \ga_{rs} T_I + 1,~~~~~~~~V=V^a \ga_a,~~~~~~~~R=\unquarto R^{ab} \ga_{ab} + F^I T_I
\en
and taking the traces on gamma matrices (see relevant formulas in Appendix B) yields the explicit first order corrections to the Einstein + Yang-Mills action. Many terms vanish because $Tr(T_I)=0$, and we find:
 \eq
  S^1_{fVRV}=  \unmezzo \theta^{AB} \int [ F^I_{AB} f^J_{rs} V^c \we F^L \we V^d -  \unmezzo f_{rs}^I V^c \we F^J_A \we F^L_B \we V^d ]\epsi_{rscd}  Tr(i \{T_I,T_J \}T_L) 
  \en
 \eqa
  & & S^1_{ffVVVV}=- {\theta^{AB} \over 12} \int  F^K_{AB} f^I_{rs} f^J_{rs} Tr(iT_K \{ T_I,T_J \}) \\
  & & S^1_{VVVV} = 0
   \ena

 Note that the first order correction $S^1$ to the NC action is proportional
to cubic powers of the gauge field strength, and vanishes
 in absence of gauge fields: this agrees with the results of
 ref. \cite{AC3}, where the first order correction to the NC action
 for vielbein gravity coupled to fermions is found to be zero. 

Moreover for all simple Lie groups $G$, except $SU(N)$ ($N \neq 2$), 
the completely symmetric tensor $D_{IJK}\equiv
Tr(T_I\{T_J,T_K\}$ vanishes.  Hence the first order in
$\theta$ correction to the NC action can be nonzero only for 
 $SU(N)$ ($N \neq 2$). The $U(1)$ case (electromagnetism coupled to gravity)
also gives a vanishing first order correction, since we have taken the $U(1)$ generator $T$ to satisfy $Tr(T)=0$.

The situation is
different for $G=SU(N)$. In this case 
$D_{IJK}$ can be different from zero  (for nonreal representations).

\sect{Conclusions and outlook}

 We have developed a systematic method to obtain deformations of classical geometric theories 
with local tangent space and gauge symmetries, based on twisted noncommutative geometry. Here we have constructed an extension of the geometrical action for the Einstein-Hilbert + Yang-Mills system, dictated by the
twisted $\star$-product of the NC action.

 Upon use of a generalization of the Seiberg-Witten map, the NC action
 has been re-expressed in terms of only the original (classical) fields: the vierbein, the spin connection and the gauge fields. The first order correction to the classical action is given explicitly, in a manifest gauge-invariant form, and is proportional to cubic powers of the gauge field strength and to the anomaly tensor $D_{IJK}$.
 
Thus NC vierbein gravity, introduced in ref. \cite{AC1}, can be consistently coupled to NC fermions
\cite{AC1} and to NC gauge fields. The coupling to NC scalar fields can be constructed via the same mechanism (first-order action) used for gauge fields \cite{AC4}. Moreover, the scalar fields can be related to the commuting vector fields $X_A$ defining the $\star$-product: then the vector fields $X_A$ become dynamical, and we obtain NC vierbein gravity with
dynamical noncommutativity \cite{AC4}, in the spirit of ref. \cite{ACD}.

\appendix

\sect{Twist differential geometry}

The noncommutative deformation of gravity considered here and in ref.s  \cite{AC1,AC2,AC3}
relies on the existence (in the deformation quantization context, see
for ex \cite{book} ) of an associative $\star$-product between
functions and more generally an associative $\westar$ exterior product between forms that
satisfies the following properties:
\sk
\noi
$\bullet~~$ \noi Compatibility with the undeformed exterior differential:
\eq
d(\tau\wedge_\star \tau')=d(\tau)\wedge_\star \tau'=\tau\wedge_\star
d\tau'
\en
$\bullet~~$ Compatibility with the undeformed integral (graded cyclicity property):
        \eq
       \int \tau \westar \tau' =  (-1)^{deg(\tau) deg(\tau')}\int \tau' \westar \tau\label{cycltt'}
       \en
      \noi with $deg(\tau) + deg(\tau')=$D=dimension of the spacetime
      manifold, and where here $\tau$ and $\tau'$ have compact support
      (otherwise stated we require (\ref{cycltt'}) to hold up to
      boundary terms).
\sk
\noi $\bullet~~$ Compatibility with the undeformed complex conjugation:
\eq
       (\tau \westar \tau')^* =   (-1)^{deg(\tau) deg(\tau')} \tau'^* \westar \tau^*
\en
{}Following \cite{AC1}  we describe here a (quite wide) class of twists whose   $\star$-products
 have all these properties.
As a particular case we
have the Groenewold-Moyal $\star$-product
\begin{equation}
f\star g = \mu \big{\{} e^{\frac{i}{2}\theta^{\rho\sigma}\partial_\rho \otimes\partial_\sigma}
f\otimes g \big{\}} , \label{MWstar}
\end{equation}
where the map $\mu$  is the usual pointwise
multiplication: $\mu (f \otimes g)= fg$, and $\theta^{\rho\sigma}$ is a constant
antisymmetric matrix.

\sk

\noi{\bf Abelian Twist}
\sk
\noi Let $\Xi$ be the linear space of smooth vector fields on a smooth manifold $M$, and $U\Xi$ its
universal enveloping algebra. A twist  ${\cal F} \in U\Xi \otimes U\Xi$
defines the associative $\star$-product
\begin{eqnarray}
f\star g &=& \mu \big{\{} {\cal F}^{-1} f\otimes g \big{\}}
\end{eqnarray}
\noi  where the map $\mu$  is the usual pointwise
multiplication: $\mu (f \otimes g)= fg$. The product associativity relies on the defining properties of the twist \cite{Wessgroup,book}.

   \noi Explicit examples of twist are provided by the so-called abelian twists:
\eq
{\cal F}^{-1}= e^{\frac{i}{2}\theta^{AB}X_A \otimes X_B} \label{Abeliantwist}
\en
where $\{X_A\}$ is a set of mutually commuting vector fields globally
defined on the manifold, and $\theta^{AB}$ is a constant
antisymmetric matrix. The corresponding $\star$-product is in general
position dependent because the vector fields $X_A$ are in general
$x$-dependent. In the special case that there exists a
global coordinate system on the manifold we can consider the
vector fields $X_A={\partial \over \partial x^A}$. In this instance we have
the Moyal twist, cf. (\ref{MWstar}):
  \eq
   {\cal F}^{-1}=  e^{\frac{i}{2}\theta^{\rho\sigma}\partial_\rho \otimes\partial_\sigma} \label{Mtwist}
   \en

  \noi {\bf Deformed exterior product}
    \sk

   \noi For abelian twists (\ref{Abeliantwist}), the deformed exterior product between forms is defined as
   \eqa
   & & \tau \westar \tau' \equiv \sum_{n=0}^\infty \left({i \over 2}\right)^n \theta^{A_1B_1} \cdots \theta^{A_nB_n}
   (\ell_{X_{A_1}} \cdots \ell_{X_{A_n}} \tau) \we  (\ell_{X_{B_1}} \cdots \ell_{X_{B_n}} \tau')  \nonumber \\
  & & ~~ = \tau \we \tau' + {i \over 2} \theta^{AB} (\ell_{X_A} \tau) \we (\ell_{X_B} \tau') + {1 \over 2!}  {\left( i \over 2 \right)^2} \theta^{A_1B_1} \theta^{A_2B_2}  (\ell_{X_{A_1}} \ell_{X_{A_2}} \tau) \we
 (\ell_{X_{B_1}} \ell_{X_{B_2}} \tau') + \cdots \nonumber 
  \label{defwestar}
  \ena
       \noi where the commuting tangent vectors $X_A$ act on forms via the Lie derivatives
       ${\ell}_{X_A} $. 
     This product is associative, and the above formula holds also for $\tau$ or $\tau'$ being a $0$-form (i.e. a function). 
 
\sk
\noi {\bf Exterior derivative}
        \sk
         \noi The exterior derivative satisfies the usual (graded) Leibniz rule,
         since it commutes with the Lie derivative:
        \eqa
        & & d (f \star g) = df \star g + f \star dg \\
        & & d(\tau \westar \tau') = d\tau \westar \tau'  + (-1)^{deg(\tau)} ~\tau \westar d\tau'
        \ena

\sk

       \noi {\bf Integration: graded cyclicity} \nopagebreak
        \sk
        \noi If we consider an abelian twist (\ref{Abeliantwist})
        given by globally defined commuting vector fields $X_A$,
        then the usual integral is cyclic under the $\star$-exterior
        products of forms, i.e., up to boundary terms,
        \eq
       \int \tau \westar \tau' =  (-1)^{deg(\tau) deg(\tau')}\int \tau' \westar \tau
       \en
      \noi with $deg(\tau) + deg(\tau')\!=\,$D$\,$= dimension of the spacetime
      manifold. In fact we have, up to boundary terms,
\eq       \int \tau \westar \tau' =    \int \tau \wedge \tau'=
(-1)^{deg(\tau) deg(\tau')}\int \tau' \wedge \tau=
(-1)^{deg(\tau) deg(\tau')}\int \tau' \westar \tau
\en
For example at first order in $\theta$,
\eq
\int \tau \westar \tau' =    \int \tau \wedge \tau'-{i\over
 2}\theta^{AB}\int{\cal L}_{X_A}(\tau\wedge {\cal L}_{X_B}\tau')
=
\int \tau \wedge \tau'-{i\over
 2}\theta^{AB}\int d {i}_{X_A}(\tau\wedge {\cal L}_{X_B}\tau')
\en
where we used the Cartan formula ${\cal L}_{X_A}=di_{X_A}+i_{X_A}d$.
\sk       \noi {\bf Complex conjugation}
    \sk
        \noi If we choose real fields $X_A$ in the definition of the
        twist (\ref{Abeliantwist}),  it is immediate to verify that:
        \eq
        (f \star g)^* = g^* \star f^*\label{starfg*}
        \en
        \eq
        (\tau \westar \tau')^* =   (-1)^{deg(\tau) deg(\tau')} \tau'^* \westar \tau^*\label{startt*}
        \en
        since sending $i$ into $-i$ in the twist (\ref{Mtwist}) amounts to send $\theta^{AB}$ into
        $-\theta^{AB} = \theta^{BA}$, i.e. to exchange the
        order of the factors in the $\star$-product.

\sect{Gamma matrices in $D=4$}

We summarize in this Appendix our gamma matrix conventions in $D=4$.

\eqa
& & \eta_{ab} =(1,-1,-1,-1),~~~\{\ga_a,\ga_b\}=2 \eta_{ab},~~~[\ga_a,\ga_b]=2 \ga_{ab}, \\
& & \ga_5 \equiv i \ga_0\ga_1\ga_2\ga_3,~~~\ga_5 \ga_5 = 1,~~~\epsi_{0123} = - \epsi^{0123}=1, \\
& & \ga_a^\dagger = \ga_0 \ga_a \ga_0, ~~~\ga_5^\dagger = \ga_5 \\
& & \ga_a^T = - C \ga_a C^{-1},~~~\ga_5^T = C \ga_5 C^{-1}, ~~~C^2 =-1,~~~C^\dagger=C^T =-C
\ena

\subsection{Useful identities}

\eqa
 & &\ga_a\ga_b= \ga_{ab}+\eta_{ab}\\
 & & \ga_{ab} \ga_5 = {i \over 2} \epsilon_{abcd} \ga^{cd}\\
 & &\ga_{ab} \ga_c=\eta_{bc} \ga_a - \eta_{ac} \ga_b -i \epsi_{abcd}\ga_5 \ga^d\\
 & &\ga_c \ga_{ab} = \eta_{ac} \ga_b - \eta_{bc} \ga_a -i \epsi_{abcd}\ga_5 \ga^d\\
 & &\ga_a\ga_b\ga_c= \eta_{ab}\ga_c + \eta_{bc} \ga_a - \eta_{ac} \ga_b -i \epsi_{abcd}\ga_5 \ga^d\\
 & &\ga^{ab} \ga_{cd} = -i \epsi^{ab}_{~~cd}\ga_5 - 4 \de^{[a}_{[c} \ga^{b]}_{~~d]} - 2 \de^{ab}_{cd}\\
& & Tr(\ga_a \ga^{bc} \ga_d)= 8~ \de^{bc}_{ad} \\
& & Tr(\ga_5 \ga_a \ga_{bc} \ga_d) = -4 ~\epsi_{abcd} \\
& & Tr(\ga^{rs} \ga_a \ga_{bc} \ga_d)=4(-2 \de^{rs}_{cd} \eta_{ab} + 2 \de^{rs}_{bd} \eta_{ac} - 3! \de^{rse}_{abc} \eta_{ed}) \\
& & Tr(\ga_5 \ga^{rs} \ga_a \ga_{bc} \ga_d)=
4(-i \eta_{ab} \epsi^{rs}_{~~cd} + i \eta_{ac} \epsi^{rs}_{~~bd} + 2i \epsi_{abc}^{~~~e} \de^{rs}_{ed})
 \ena
\sk
\noi where
$\delta^{ab}_{cd} \equiv \frac{1}{2}(\delta^a_c\delta^b_d-\delta^b_c\delta^a_d)$, $\delta^{rse}_{abc} \equiv  {1 \over 3!} (\de^r_a \de^s_b \de^e_c$ + 5 terms), 
and indices antisymmetrization in square brackets has total weight $1$.

 \subsection{Charge conjugation and Majorana condition}

\eqa
 & &   {\rm Dirac~ conjugate~~} \psibar \equiv \psi^\dagger
 \ga_0\\
 & &  {\rm Charge~ conjugate~spinor~~} \psi^C = C (\psibar)^T  \\
 & & {\rm Majorana~ spinor~~} \psi^C = \psi~~\Rightarrow \psibar =
 \psi^T C
 \ena


\begin{thebibliography}{99}

\bibitem{SW}
N.~Seiberg and E.~Witten, {\it String theory and noncommutative
geometry}, JHEP  {\bf 9909}, 032 (1999) [hep-th/9908142].

\bibitem{MoyalGroenewold}
J.~E.~Moyal,  {\it Quantum mechanics as a statistical theory}, 
Proc. Camb. Phil. Soc. {\bf 45} (1949) 99 ;
H. Groenewold, Physica {\bf 12} (1946) 405.

\bibitem{Weyl}
  H.~Weyl,
  ``Quantum mechanics and group theory,''
  Z.\ Phys.\  {\bf 46}, 1 (1927).

 \bibitem{AC1}
  P.~Aschieri and L.~Castellani,
  ``Noncommutative D=4 gravity coupled to fermions,''
  JHEP {\bf 0906} (2009) 086
  [arXiv:0902.3817 [hep-th]].

\bibitem{AC2}
  P. Aschieri and L. Castellani,
  ``Noncommutative supergravity in D=3 and D=4,''
  JHEP {\bf 0906} (2009) 087
  [arXiv:0902.3823 [hep-th]].

\bibitem{Madore}
  J.~Madore, S.~Schraml, P.~Schupp, J.~Wess,
  ``Gauge theory on noncommutative spaces,''
  Eur.\ Phys.\ J.\  {\bf C16 } (2000)  161-167.
  [hep-th/0001203].

\bibitem{starYM} 
  B.~Jurco, L.~Moller, S.~Schraml, P.~Schupp, J.~Wess,
  ``Construction of nonAbelian gauge theories on noncommutative spaces,''
  Eur.\ Phys.\ J.\  {\bf C21 } (2001)  383-388.
  [hep-th/0104153].

\bibitem{AJSW}
  P.~Aschieri, B.~Jurco, P.~Schupp, J.~Wess,
  ``Noncommutative GUTs, standard model and C,P,T,''
  Nucl.\ Phys.\  {\bf B651 } (2003)  45-70.
  [hep-th/0205214].

\bibitem{AC3} 
  P.~Aschieri and L.~Castellani,
  ``Noncommutative gravity coupled to fermions: second order expansion via Seiberg-Witten map,''
  arXiv:1111.4822 [hep-th].

\bibitem{ACD}
  P.~Aschieri, L.~Castellani and M.~Dimitrijevic,
  ``Dynamical noncommutativity and Noether theorem in twisted $\phi^{\star 4}$ theory,''
  Lett.\ Math.\ Phys.\  {\bf 85} (2008) 39
  [arXiv:0803.4325 [hep-th]].

\bibitem{AC4}
  P.~Aschieri and L.~Castellani,``Vielbein gravity with dynamical noncommutativity", to appear

\bibitem{CDF} 
  L.~Castellani, R.~D'Auria and P.~Fr\'e,
  ``Supergravity and superstrings: A Geometric perspective,''
  Singapore, Singapore: World Scientific (1991)

\bibitem{C92} 
  L.~Castellani,
  ``Group geometric methods in supergravity and superstring theories,''
  Int.\ J.\ Mod.\ Phys.\ A {\bf 7}, 1583 (1992).

\bibitem{Ulker}
  K.~Ulker, B.~Yapiskan,
  ``Seiberg-Witten maps to all orders,''
  Phys.\ Rev.\  {\bf D77 } (2008)  065006.
  [arXiv:0712.0506 [hep-th]].

\bibitem{Wessgroup}
P.~Aschieri, C.~Blohmann, M.~Dimitrijevi\' c, F.~Meyer, P.~Schupp
and J.~Wess, {\it A Gravity Theory on Noncommutative Spaces},
Class.\ Quant.\ Grav. {\bf 22}, 3511-3522 (2005),
[hep-th/0504183];

  \bibitem{book}
 P.~Aschieri, M. Dimitrijevic, P. Kulish, F. Lizzi, J. Wess,
 ``Noncommutative Spacetimes", Lecture Notes in Physics, vol. 774, Springer 2009.


\end{thebibliography}
\end{document}